\documentclass[showpacs,twocolumn,pre]{revtex4}
\usepackage{amsmath}
\usepackage{latexsym}
\usepackage{epsfig}
\usepackage{makeidx}
\usepackage{amsfonts}

\begin{document}

\title{Sub-diffusion in External Potential: Anomalous hiding behind Normal}
\author{Sergei Fedotov, Nickolay Korabel }
\affiliation{School of Mathematics, The University of Manchester, Manchester M13 9PL, UK}

\begin{abstract}
We propose a model of sub-diffusion in which an external force is acting
on a particle at all times not only at the moment
of jump. The implication of this assumption is the dependence of the random trapping
time on the force with the dramatic change of particles behavior
compared to the standard continuous time random walk model. Constant force
leads to the transition from non-ergodic sub-diffusion to seemingly ergodic
diffusive behavior. However, we show it remains anomalous in a sense that
the diffusion coefficient depends on the force and the anomalous exponent.
For the quadratic potential we find that the anomalous exponent defines not
only the speed of convergence but also the stationary distribution which is
different from standard Boltzmann equilibrium.
\end{abstract}



\maketitle

Recently it has become clear that anomalous diffusion measured by a
non-linear growth of the ensemble averaged mean squared displacement $%
\left\langle x^{2}\right\rangle \sim t^{\mu}$ with the anomalous exponent $%
\mu \neq 1$ is as widespread and important as normal diffusion with $\mu =1$
\cite{Kla08}. Sub-diffusion with $\mu <1$ was observed in many physical
and biological systems such as porous media \cite{Dra99}, glass-forming
systems \cite{Weeks02}, motion of single viruses in the cell \cite{Sei01},
cell membranes \cite{Sax97,Rit05}, and inside living cells \cite%
{Wei03,Gol06,Tolic04}. Many examples of sub-diffusive processes in
biological systems can be found in recent reviews \cite{Fra13,Bar12}.
Nowadays new tools are available including super-resolution light optical
microscopy techniques to deal with biological in vivo data which allows to
monitor a large number of trajectories at the single-molecule level and at
nanometer resolution \cite{Ser08,Jaq08,Nor14}. Using these techniques
it is possible to discriminate between anomalous ergodic processes where the
ensemble and time averages coincide and non-ergodic processes where ensemble
and time averages have different behavior \cite{He08,Lub08,Mer13}. Two
important observations have been made about anomalous transport in living
cells: (1) anomalous transport is usually a transient phenomenon before
transition to normal diffusion or saturation due to confined space \cite%
{Bro09,Neu08,Sax01} (2) ergodic and non-ergodic processes may coexist as it
was observed in plasma membrane \cite{Wei11}.

Several models are proposed to describe ergodic and non-ergodic anomalous
processes  such as non-ergodic continuous time random walk (CTRW) with
power-law tail waiting times, ergodic anomalous process
generated by fractal structures, fractional Brownian-Langevin
motion characterized by long correlations and time
dependent diffusion coefficient \cite{Kla08,Sok12,Bre13}. 
The standard CTRW model for
sub-diffusion of a particle in an external field $F(x)$ randomly moving
along discrete one-dimensional lattice can be described by the generalized
master equation for the probability density $p(x,t)$ to find the particle at
position $x$ at time $t$
\begin{equation}
\frac{\partial p}{\partial t}=-i(x,t)+w^{+}(x-a)i(x-a,t)+w^{-}(x+a)i(x+a,t),
\label{MM}
\end{equation}%
where $a$ is the lattice spacing and $i(x,t)$ is the total escape rate from $%
x$
\begin{equation}
i(x,t)=\frac{1}{\Gamma(1-\mu)\tau _{0}^{\mu }}\mathcal{D}_{t}^{1-\mu }p(x,t).  \label{ieq}
\end{equation}%
Here $\tau _{0}$ is a constant timescale and  $\mathcal{D}_{t}^{1-\mu }$ is
the Riemann-Liouville fractional derivative defined by
\begin{equation}
\mathcal{D}_{t}^{1-\mu }p(x,t)=\frac{1}{\Gamma (\mu )}\frac{\partial }{%
\partial t}\int_{0}^{t}\frac{p(x,\tau )}{(t-\tau )^{1-\mu }}d\tau .
\end{equation}%
The probabilities of jumping to the right $w^{+}(x)$ and to the left $w^{-}(x)$ are
\begin{equation}
w^{+}(x)=\frac{1}{2}+\beta a F(x),\quad w^{-}(x)=\frac{1}{2}-\beta a F(x).
\label{ww}
\end{equation}%
Series expansion of Eq.\ (\ref{MM}) together with Eq.\ (\ref{ieq}) and Eq.\ (\ref{ww}) 
leads to the fractional Fokker-Planck equation (FFPE) \cite{Met99,Met00}
\begin{equation}
\frac{\partial p}{\partial t}=D_{\mu }\left[ \frac{\partial ^{2}}{\partial
x^{2}}-\beta \frac{\partial }{\partial x}F(x)\right] \mathcal{D}_{t}^{1-\mu
}p,  \label{eq:FFPE}
\end{equation}%
where the generalized diffusion $D_{\mu }=a^{2}/(2\;\Gamma (1-\mu )\tau
_{0}^{\mu }).$ The stationary solution of Eq.\ (\ref{eq:FFPE}) is the Boltzmann
distribution. There exist a huge literature on this equation \cite%
{Met99,Met00} and its generalization for time dependent forces \cite%
{Hei07,Mag08,Henry10,Eule09,Sok06,Shu08,Shkilev12}.

One of the main assumptions in this literature, which is not always clearly
stated is that, as long as a random walker is trapped at a particular point $%
x$, the external force $F(x)$ does not influence the particle. It is clear
from Eq.\ (\ref{ieq}) that the escape rate $i(x,t)$ does not depend on the
external force $F(x).$ The force only acts at the moment of escape inducing a bias. The
question is how to take into account the dependence of the escape rate on $%
F(x)?$ To the author's knowledge this is still an open question. One of the
main aims of this Letter is to propose a model which deals with this
problem. We find that the dependence of escape rate on force drastically
changes the form of the master equation (\ref{MM}) and FFPE (\ref{eq:FFPE}).
We observe transient anomalous diffusion and transition from non-ergodic to
normal ergodic behavior. However, we show that this seemingly normal process
could be still anomalous masked by normal behavior. Our findings suggest
that a closer inspection of experimental results could be necessary in order
to discriminate between normal and anomalous processes.

\textit{Model.---} We consider a random particle moving on a one dimensional
lattice under assumption that an external force acts on a particle at all
times not only at the moment of jump as in Eq.\ (\ref{MM}). The implication
of this assumption is the dependence of the random trapping time on the
external force (not just jumping probabilities as in (\ref{ww})). Some
discussion of situation when the external force influence the rates and
jumps can be found in \cite{Hei07}. The main physical idea behind our model
is that there exists two independent mechanism of escaping from the point $x$
with two different random residence times. The first mechanism is due to
external force with the escape rate proportional to $F(x).$ The second one
is the sub-diffusive mechanism involving the rate inversely proportional to
the residence time. The latter generates the power law waiting time
distribution with the infinite first moment.

Regarding the first mechanism, we define the jump process from the point $x$
as follows. We assume that the rate of jump to the
right $\mathbb{T}_{x}^{+}$ from $x$ to $x+a$ is $\nu aF(x)$ when $F(x)\geq 0$ and the rate of jump to the left
$\mathbb{T}_{x}^{-}$ from $x$ to $x-a$ \ is $-\nu
aF(x)$ when $F(x)\leq 0.$ For this jump model the random waiting time $T_{F}$
at the point $x$ is defined by the exponential survival probability $\Psi
_{F}(x,\tau )$ involving the external force $F(x)$
\begin{equation}
\Psi _{F}(x,\tau )=\Pr \left\{ T_{F}>\tau \right\} =\exp \left( -\nu
a|F(x)|\tau \right).
\end{equation}%
where $\nu $ is the intensity of jumps due to force field. For example, one
can think of the escape rate $\mathbb{T}_{x}^{+}$ that is defined in
terms of the potential field $U(x)$ that is $\mathbb{T}_{x}^{+}=-\nu \left[
U(x+a)-U\left( x\right) \right] >0,$ there $F(x)=-U^{\prime }(x)+o(a^{2})$
for $U^{\prime }(x)\leq 0.$ The second mechanism involves the sub-diffusive
random walk with the escape rate $\lambda (x,\tau )$ from the point $x$,
which is inversely proportional to the residence time $\tau.$ In this case
the random waiting time $T_{\lambda }$ at the point $x$ is defined by the
survival probability
\begin{equation}
\Psi _{\lambda }(x,\tau )=\Pr \left\{ T_{\lambda }>\tau \right\} =\exp
\left( -\int_{0}^{\tau }\lambda (x,s)ds\right) .  \label{FFFF}
\end{equation}%
The question now is how to implement the jumping process due to external
force into the sub-diffusive random walk scheme? When the random walker
makes a jump to the point $x$, it spends some random time (residence time)
before making another jump to $x+a$ or $x-a$. Let us denote this residence
time $T_{x}$. The key point of our model is that we define this residence
time as the minimum of two: $T_{\lambda }$ and $T_{F}$
\begin{equation}
T_{x}=\min \left( T_{\lambda },T_{F}\right) .  \label{min}
\end{equation}%
For the anomalous sub-diffusive case this model could lead to the drastic
change in the form of the fractional master equation. The main reason for
this is that the external force $F(x)$ plays the role of tempering factor
preventing the random walker to be anomalously trapped at point $x$.
\begin{figure}[t]
\centerline{
\psfig{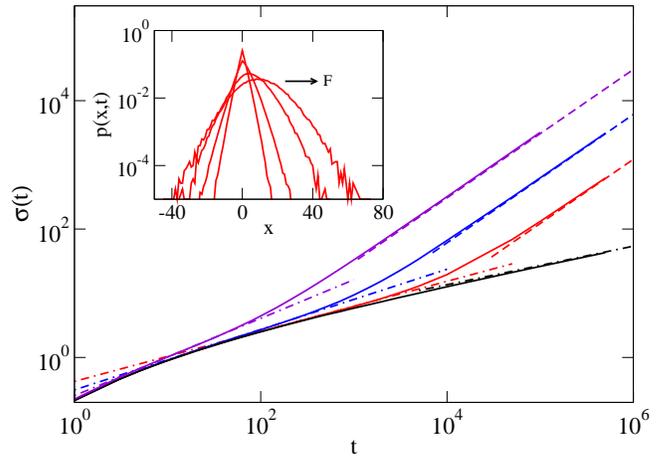}
}
\caption{Variance $\protect\sigma (t)$ of ensemble calculated with $\protect%
\mu =0.3$ and $p(x,0)=\protect\delta (x)$. In all simulations we use $\nu=1$. 
For $F=0$ (lowest curve) the
variance grows as $D_{\protect\mu }t^{\protect\mu }$ (dashed-dotted line) in
the limit $t\rightarrow \infty $. Constant force $F=0.0001$, $F=0.001$ and $%
F=0.01$ (curves from bottom to top on the RHS of the figure)
leads to the transition from sub-diffusive behavior for short times to
normal diffusion in the long time limit, $\protect\sigma \rightarrow 2D_{F}t$
(dashed lines), with $D_{F}$ given by Eq.\ (\protect\ref{D_F}). Intermediate
asymptotic of the variance is fitted by the power law (dashed-dotted lines, see
the text). The inset shows transition of densities from sub-diffusive form
for short times to the Gaussian shape for long times caused by the constant
force $F=0.0001$. Densities were calculated at $t=10^{3}$, $10^{4}$, $5\cdot
10^{4}$ and $10^{5}$.}
\label{FIG1}
\end{figure}
Because of the independence of two mechanisms, in our model the rate of jump
$\mathbb{T}_{x}^{+}$ to the right from $x$ to $x+a$ and the rate of jump $%
\mathbb{T}_{x}^{-}$ to the left from $x$ to $x-a$ can be written as the sum%
\begin{equation}
\mathbb{T}_{x}^{+}=%
\begin{cases}
\omega ^{+}(x)\lambda (x,\tau )+\nu aF(x), & F(x)\geq 0, \\
\omega ^{-}(x)\lambda (x,\tau ),\quad  & F(x)<0%
\end{cases}
\label{T+}
\end{equation}%
and%
\begin{equation}
\mathbb{T}_{x}^{-}=%
\begin{cases}
\omega ^{+}(x)\lambda (x,\tau ), & F(x)\geq 0, \\
\omega ^{-}(x)\lambda (x,\tau )-\nu aF(x),\quad  & F(x)<0.%
\end{cases}
\label{T-}
\end{equation}%
Although it is straightforward to consider general $\omega ^{+}(x)$, $\omega
^{-}(x)$, for simplicity in what follows we consider $\omega ^{+}(x)=\omega
^{-}(x)=1/2$. In our model the asymmetry of random walk occur only from the
force dependent rate. Let us explain the main idea of Eqs.\ (\ref{T+}) and (%
\ref{T-}). The external force $F(x)\geq 0$ increases the sub-diffusive rate
of jumps to the right $\lambda (x,\tau )/2$ and does not change the
sub-diffusive rate of jumps to the left. The essential property of Eqs.\ (%
\ref{T+}) and (\ref{T-}) is that the rate $\lambda (x,\tau )$ depends on the
residence time variable $\tau $. This dependence makes any model involving
the probability density $p(x,t)$ non-Markovian. For the Markov case with $%
F(x)=0,$ $\lambda ^{-1}(x)$ has a meaning of the mean residence time at the
point $x$. When the parameter $\nu =0$ and the rates are $\mathbb{T}%
_{x}^{+}=w^{+}(x)\lambda (x,\tau ),$ $\mathbb{T}_{x}^{-}=w^{-}(x)\lambda
(x,\tau )$, we obtain the standard fractional Fokker-Planck equation (\ref%
{eq:FFPE}). Notice that Eq.\ (\ref{min}) is consistent with the expression
for the effective escape rate $\mathbb{T}_{x}^{+}+\mathbb{T}_{x}^{-}$ as a
sum of two rates $\lambda (x,\tau )+\nu a|F(x)|$. Similar situation has been
considered in \cite{Fed13}.

After incorporation of the force dependent escape rates we can obtain
generalized master equation (see Supplementary Materials for the
derivation). By expanding the RHS of the master equation to the second order
in jump size $a$, we get a fractional diffusion equation
\begin{equation}
\frac{\partial p}{\partial t}=\frac{\partial ^{2}}{\partial x^{2}}\left[
D_{\mu }e^{-\nu a|F(x)|t}\mathcal{D}_{t}^{1-\mu }\left[ p(x,t)e^{\nu
a|F(x)|t}\right] \right] -  \label{main_eq}
\end{equation}%
\begin{equation*}
-a^{2} \nu \frac{\partial }{\partial x}\left[ F(x)p(x,t)\right] .
\end{equation*}
This equation is fundamentally different from the classical FFPE (\ref%
{eq:FFPE}) because it involves the external force in both terms on the right
hand side. One can see that the force $F(x)$ not only determines the
advection term as in Eq.\ (\ref{eq:FFPE}), but also plays the role of
tempering parameter through the factor $e^{\nu a|F(x)|t}$. Similar factor
occurs in sub-diffusive equation with the death or evanescent process \cite{Abad10,Fal13}. 
However, here we consider the system with constant total number of particle.

The stationary solution $p_{st}(x)$ of Eq.\ (\ref{main_eq}) obeys the
standard equation%
\begin{equation}
-a^{2}\nu F(x)p_{st}(x)+\frac{d}{dx}\left[ D_{F}(x)p_{st}(x)\right] =0.
\label{stat_eq}
\end{equation}%
(see a supplement material for details). Interesting property of this
equation is that the effective diffusion constant $D_{F}(x)$ depends on the
external force and anomalous exponent
\begin{equation}
D_{F}(x)=D_{\mu }\left( \nu a|F(x)|\right) ^{1-\mu }.  \label{D_F}
\end{equation}%
This fact implies that the Boltzmann distribution is no longer stationary
solution of (\ref{stat_eq}). For the quadratic potential $U(x)=\kappa x^{2}/2
$ with $F(x)=kx,$ we find that for large $x$ the stationary density $%
p_{st}(x)$ has the form
\begin{equation}
p_{st}(x)\sim \exp (-A|x|^{1+\mu }),  \label{stat}
\end{equation}%
where $A>0$ is a constant. One can see that the form of stationary density
is determined by the anomalous exponent $\mu $. In this case the particles
spread further compared to the Boltzmann case. The reason is the dependence
of the effective diffusion constant $D_{F}(x)$ on force $F(x).$ Note that
for the sub-diffusive fractional Fokker-Planck equation (\ref{eq:FFPE}) the
anomalous exponent only determines the slow power law relaxation rate, while
the stationary density converges to Boltzmann equilibrium which does not
depend on $\mu .$

\textit{Numerical simulations.---} We consider two particular cases: \ (1)
constant force $F$ corresponding to the linear potential and (2) the
quadratic potential $U(x)=\kappa x^{2}/2$ both in the infinite domain. We
concentrate on the behavior of the density function $p(x,t)$, the mean $%
\left\langle x(t)\right\rangle $ and the variance $\sigma (t)=\left\langle
x^{2}\right\rangle -\left\langle x\right\rangle ^{2}$ calculated using an
ensemble of trajectories from the initial distribution $p(x,0)=\delta (x)$.
We also calculate the time averaged variance of a single trajectory of
length $T$, $\sigma _{T}(\Delta ,T)=\delta ^{2}(\Delta ,T)-(\delta (\Delta
,T))^{2}$, where $\delta ^{n}(\Delta ,T)=\int_{0}^{T-\Delta }(x(t+\Delta
)-x(t))^{n}dt/(T-\Delta )$, $n=1,2$. This quantity become a standard tool to
assess ergodic properties of a system been equivalent to its ensemble
averaged counterpart only for ergodic case.

\begin{figure}[t]
\centerline{
\psfig{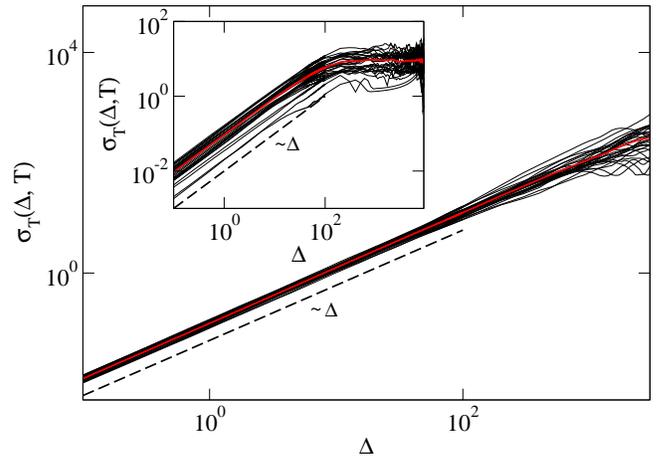}
}
\caption{Time averaged variance $\protect\sigma _{T}(\Delta ,T)$ calculated
for $30$ individual trajectories of the length $t=10^{4}$ (each curve corresponds to a single
trajectory) with $\protect\mu =0.7$. The minor
scatter between trajectories reflects the ergodic behavior under the action
of constant external force (main figure). Contrast this with the behavior of
the time averaged variance in quadratic potential (the inset). The scatter
between individual trajectories indicates that the system is non-ergodic in
this case. The red (bold solid) lines represent average over $30$
trajectories. }
\label{FIG2}
\end{figure}

When the external force $F$ is constant, we observe the transition from
sub-diffusion at short times to seemingly normal diffusion at long times.
The density function changes from the distinct sub-diffusive shape for short
times to the Gaussian shape propagator at longer times (see the inset of
figure \ref{FIG1}). The average position of the ensemble behaves as $\left<x(t)\right>=F t$. 
The ensemble averaged variance $\sigma(t)$ grows as a
power law for short times, $\sigma(t) \sim t^{\eta}$, and transition to a
normal diffusive linear growth $\sigma (t)\sim 2D_{F}t$ for longer times.
However, in this case the diffusion coefficient $D_{F}$ depends on the force
$F$ and anomalous exponent $\mu $. We conclude that although the variance $%
\sigma (t)$ is linearly proportional to time, this dependence reveals the
anomalous nature of the process even in the limit $t\rightarrow \infty $.
Numerical calculations confirms the analytical result for the diffusion
coefficient Eq.\ (\ref{D_F}) (see figure \ref{FIG1}). Second observation is
that the power law behavior at short times involves the exponent $\eta
(F)>\mu $ which depends on force $F$. For $\mu =0.3$ they are estimated to be $\eta
\approx 0.39$ for $F=0.0001$, $\eta \approx 0.47$ for $F=0.001$ and $\eta
\approx 0.6$ for $F=0.01$. This can be interpreted as an enhancement of
sub-diffusion coursed by the constant force. Such enhancement should be
taken into account in the analysis of biological experiments where
sub-diffusion usually appears as transient before the transition to the
normal diffusion \cite{Fra13}. For the large value of $F$ the exponent $\eta
$ tends to one while in the small force limit $\eta \rightarrow \mu $. The
time averaged variance calculated for constant force grows linearly $\sigma
_{T}(\Delta ,T)\sim \Delta $ and shows minor scatter between single
trajectories (figure \ref{FIG2}). After averaging over different
trajectories, it grows with the coefficient $2 D_F$ which is equal to the
ensemble average value. This shows that the non-ergodic sub-diffusive system
becomes an ergodic one.
\begin{figure}[t]
\centerline{
\psfig{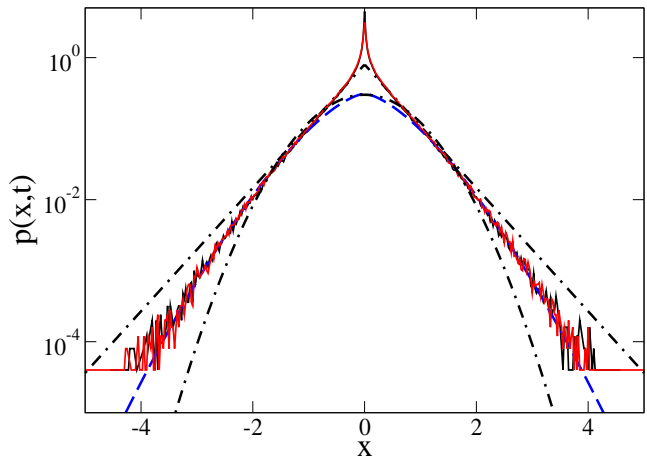}
}
\caption{Density $p(x,t)$ in the quadratic potential $U(x)=kx^{2}$, $k=0.001$
calculated with the anomalous exponent $\protect\mu =0.5$ at time $t=10^{5}$
and $t=10^{6}$. Two densities overlap indicating convergence to stationary
solution $p_{st}$. Clearly $p_{st}$ is non-Boltzmann and is well described
(accept for the central part) by the long-wave asymptotic Eq.\ (\protect\ref%
{stat_eq}) shown by the dashed line. To distinguish the form of the
stationary solution $\exp (-A|x|^{1+\protect\mu })$, we show the Boltzmann
equilibrium $\exp (-B x^2)$ and the function $\exp (-C|x|)$
(dashed-dotted curves) to guide the eye ($A, B, C$ are positive constants). Note that the central part of $%
p_{st}$ has distinct cusp at $x=0$ where the force vanishes. }
\label{FIG3}
\end{figure}

Now we consider the quadratic potential $U(x)=\kappa x^{2}/2.$ The system
becomes again non-ergodic despite the tempering affect of the force. To
confirm this we calculate the time averaged variance (inset of figure \ref%
{FIG2}). As expected it shows large fluctuations among different
trajectories typical for non-ergodic systems. Note that even with this
typical behavior, it can be easily distinguished in experiments since in our
case the mean of the time averaged variance converges to a constant, while
for standard CTRW in a bounded region it grows as a power of the anomalous
exponent, $\left\langle \sigma_{T}(\Delta )\right\rangle \sim \Delta ^{1-\mu
}$. Regarding the shape of the stationary density, numerical simulations are
in good agreement with analytical results Eq.\ (\ref{stat}) (see figure \ref%
{FIG3}).

\textit{Summary.---} In this Letter we have presented a model of anomalous
sub-diffusive transport in which the force acts on the particle at all times
not only at the moment of jump. This leads to the dependence of jumping rate
on the force with the dramatic change of particles behavior compared to the
standard CTRW model.  We have derived a new type of fractional diffusion
equation which is fundamentally different from the classical fractional
Fokker-Planck equation. In our model the force $F(x)$ not only appears in
the drift term as in Eq.\ (\ref{eq:FFPE}), but also determines the structure
of the diffusion term controlling the spread of particles. The constant
external force leads to the natural tempering of the broad waiting time
distribution and, as a result, to the transition to a seemingly normal
diffusion (linear growth of the mean squared displacement) and equivalence
of the time and ensemble averages. However, this may lead to a wrong conclusion
in analyzes of experimental results on transient sub-diffusion \cite{Fra13}
that the process is normal for large times. We have found that contrary to
normal diffusion process in the external force field, the diffusion
coefficient depends on the force and anomalous exponent. This fact implies
that the Boltzmann distribution is no longer stationary solution. External
perturbations and noise fluctuations are not separable which reflects the
non-Markovian nature of the process even for large times.

Our results would be possible to test in experiments, for example, by
considering a bead which is moving sub-diffusively in an actin network. The
motion of such a beat can be described by a random walk type of dynamics
\cite{Wong04}. Force-measurements could be realized by using optical trap
and tweezers which are the nano-tools capable of performing such
measurements on individual molecules and organelles within the living cell
\cite{Nor14}. When the force is constant the dependence of the measures
diffusion coefficient on the strength of the force would reveal the
predicted power law behavior $F^{1-\mu }$. For quadratic potential it could
be possible to retrieve the form of the stationary profile (\ref{stat}) with
the slow decay compared to Boltzmann distribution for large $x$.\

\section*{Acknowledgements}

SF and NK acknowledge the support of the EPSRC Grant EP/J019526/1.


\newpage

\section*{SUPPLEMENTARY MATERIALS}

To derive Eq.\ (\ref{main_eq}) we use the structured probability density
function $\xi (x,t,\tau )$ with the residence time $\tau $ as auxiliary
variable. This density gives the probability that the particle position $%
X(t) $ at time $t$ is at the point $x$ and its random residence time $T_{x}$ at
point $x$ is in the interval $(\tau ,\tau +d\tau ).$ The density $\xi
(x,t,\tau )$ obeys the balance equation
\begin{equation}
\frac{\partial \xi }{\partial t}+\frac{\partial \xi }{\partial \tau }%
=-\left( \mathbb{T}_{x}^{+}(x,\tau )+\mathbb{T}_{x}^{-}(x,\tau )\right) \xi .
\label{basic}
\end{equation}%
We consider only the case when the residence time of random walker at $t=0$
is equal to zero, so the initial condition is
\begin{equation}
\xi (x,0,\tau )=p_{0}(x)\delta (\tau ),  \label{initial}
\end{equation}%
where $p_{0}(x)$ is the initial density. The boundary condition in terms of
residence time variable ($\tau =0)$ can be written as \cite{Cox}
\begin{eqnarray}
\xi (x,t,0) &=&\int_{0}^{t}\mathbb{T}_{x}^{+}(x-a,\tau )\xi (x-a,t,\tau
)d\tau +  \notag \\
&&\int_{0}^{t}\mathbb{T}_{x}^{-}(x+a,\tau )\xi (x+a,t,\tau )d\tau .
\label{arr1}
\end{eqnarray}%
We solve (\ref{basic}) by the method of characteristics for $\tau <t$
\begin{equation}
\xi (x,t,\tau )=j\left( x,t-\tau \right) \Psi_{\lambda} (x,\tau )e^{-\Phi(x) \tau
},\quad \tau <t,  \label{je}
\end{equation}%
where 
\begin{equation}
\Phi(x) = \nu a |F(x)|.
\end{equation}%
The solution Eq.\ (\ref{je}) is written in terms of the integral arrival rate $%
j(x,t)=\xi (x,t,0)$ and in terms of the survival function Eq. (\ref{FFFF})
\begin{equation}
\Psi_{\lambda} (x,\tau ) = e^{-\int_{0}^{\tau } \lambda (x,s) ds}.
\end{equation}%
Our purpose now is to derive the master equation for the probability density%
\begin{equation}
p(x,t)=\int_{0}^{t^{+}}\xi (x,t,\tau )d\tau .  \label{denG}
\end{equation}%
Let us introduce the integral escape rate to the right $i^{+}(x,t)$ and the
integral escape rate to the left $i^{-}(x,t)$ as%
\begin{equation}
i^{\pm }(x,t)=\int_{0}^{t^{+}} \omega^{\pm}(x) \lambda(x,\tau )\xi (x,t,\tau )d\tau .
\label{i1}
\end{equation}%
Note that the integration with respect to the residence time $\tau $ in (\ref%
{denG}) and (\ref{i1}) involves the upper limit $\tau =t,$ where we have a
singularity due to the initial condition (\ref{initial}). Then the boundary
conditions (\ref{arr1}) can be written in a simple form:
\begin{eqnarray}
j(x,t) &=&i^{+}(x-a,t)+i^{-}(x+a,t)  \notag \\
&&+%
\begin{cases}
\nu aF(x-a)p(x-a,t), & F\geq 0, \\
-\nu aF(x+a)p(x+a,t),\quad & F<0.%
\end{cases}
\label{jj}
\end{eqnarray}%
It follows from (\ref{initial}), (\ref{je}) and (\ref{i1}) that
\begin{eqnarray}
i^{\pm }(x,t) &=&\int_{0}^{t}\psi ^{\pm }(x,\tau )j(x,t-\tau )e^{-\Phi(x)
\tau }d\tau  \notag \\
&&+\psi ^{\pm }(x,t)p_{0}(x)e^{-\Phi(x) t},  \label{i55}
\end{eqnarray}%
where $\psi ^{+}(x,\tau )=\omega^{+}(x) \lambda(x,\tau ) \Psi_{\lambda} (x,\tau )$ and 
$\psi^{-}(x,\tau )=\omega^{-}(x) \lambda(x,\tau )\Psi_{\lambda} (x,\tau ).$ Substitution of (\ref%
{initial}) and (\ref{je}) to (\ref{denG}), gives
\begin{eqnarray}
p(x,t) &=&\int_{0}^{t}\Psi_{\lambda} (x,\tau )j(x,t-\tau )e^{-\Phi(x) \tau }d\tau
\notag \\
&&+\Psi_{\lambda} (x,t)p_{0}(x)e^{-\Phi(x) t}.  \label{p11}
\end{eqnarray}%
The balance equation for probability density $p(x,t)$ can be written as%
\begin{equation}
\frac{\partial p}{\partial t}=-i^{+}(x,t)-i^{-}(x,t)+j(x,t)-\Phi(x) p(x,t).
\label{balan}
\end{equation}%
Let us find a closed equation for $p(x,t)$ by expressing integral rates $%
i^{\pm }(x,t)$ and $j(x,t)$ in terms of the density $p(x,t).$ We apply the
Laplace transform $\hat{f}(s)=\int_{0}^{\infty }f (\tau )e^{-s\tau
}d\tau $ to (\ref{i55}), and (\ref{p11}), and obtain
\begin{equation}
\hat{i}^{\pm }(x,s)=\frac{\hat{\psi}^{\pm }(x,s+\Phi(x))}{\hat{\Psi}%
(x,s+\Phi(x))}\hat{p}(x,s),  \label{new55}
\end{equation}%
which after the inversion of the Laplace transform and using the shift
theorem gives
\begin{equation}
i^{\pm }(x,t)=\int_{0}^{t}K^{\pm }(x,t-\tau )e^{-\Phi(x) (t-\tau)} p(x,\tau)
d\tau.
\end{equation}%
The memory kernels $K^{+}(x,t)$ and $K^{-}(x,t)$ are defined by Laplace
transforms
\begin{equation}
\hat{K}^{\pm }\left( x,s\right) =\hat{\psi}^{\pm }(x,s)/\hat{\Psi}_{\lambda}\left(
x,s\right).  \label{new5}
\end{equation}%
Now we consider the sub-diffusive case where $\lambda (\tau)$ is inversely proportional to
the residence time $\tau:$
\begin{equation}
\lambda (\tau )= \mu/(\tau _{0}+\tau),\qquad 0<\mu <1.
\label{s33}
\end{equation}%
For simplicity we consider 
\begin{equation}
\omega^{-}=\omega^{+}=1\slash2.
\end{equation}
It is straightforward to generalize to non-homogeneous systems by considering
space dependent $\lambda(x)$ and space dependent anomalous exponent $\mu(x)$, 
this case we consider elsewhere \cite{Che05,Fed13}. From Eqs.\ (\ref{FFFF}) and (\ref%
{s33}) it follows that the survival function has a power-law dependence
\begin{equation}
\Psi_{\lambda} (\tau)= \tau _{0}^{\mu} \left( \tau _{0}+\tau \right)^{-\mu}.
\end{equation}
The waiting time density functions $\psi^{\pm} (\tau )$ are
\begin{equation}
\psi^{+}(\tau)=\psi^{-}(\tau)= \mu \tau _{0}^{\mu} (\tau _{0}+\tau )^{-1-\mu}\slash2.
\label{Pareto}
\end{equation}%
Using the Tauberian theorem their Laplace transforms are $\hat{\psi}%
^{\pm}\left( s\right) \simeq (1- g s^{\mu})/2$ as $s\rightarrow 0$, where $g =
\Gamma (1-\mu )\tau _{0}^{\mu}$. From (\ref{new5}) we obtain the
Laplace transforms
\begin{equation}
\hat{K}^{+}(s) = \hat{K}^{-} (s) \simeq s^{1-\mu}\slash (2 g),\qquad s\rightarrow 0.
\end{equation}%
Therefore, the integral escape rates to the right $i^{+}$ and to the left $%
i^{-}$ in the sub-diffusive case are
\begin{equation}
i^{+}(x,t)=i^{-}(x,t) = e^{-\Phi(x) t}\mathcal{D}_{t}^{1-\mu}\left[ p(x,t)e^{\Phi(x)
t}\right] \slash (2 g).  \label{gen_i}
\end{equation}%
By introducing the total integral escape rate 
\begin{equation}
i(x,t)=i^{+}(x,t)+i^{-}(x,t),
\end{equation}%
and expanding the right-hand side of Eq.\ (\ref{balan}) to second order in jump size $a$ we obtain
the following fractional equation
\begin{equation}
\frac{\partial p}{\partial t} = - a^{2}\nu \frac{\partial }{\partial x}\left[ F(x)p(x,t)\right] + \frac{a^{2}}{2}\frac{\partial ^{2}i}{\partial x^{2}},
\label{sup_main}
\end{equation}%
which using Eq.\ (\ref{gen_i}) leads to the main equation of the paper Eq.\ (\ref{main_eq}).

Now we derive the equation for the stationary solution Eq.\ (\ref{stat_eq}). Writing the escape rate $i(x,t)$ 
in Laplace form
\begin{equation}
\hat{i}(x,s) = \frac{(s+\Phi(x))^{1-\mu}}{g} \hat{p}(x,s)
\end{equation}%
and taking the limit $s \rightarrow 0$ corresponding to $t \rightarrow \infty$, we obtain the 
stationary escape rate
\begin{equation}
i_{st}(x) = \frac{\Phi(x)^{1-\mu}}{g} p_{st}(x).
\label{i_st}
\end{equation}%
where the stationary density is defined in a standard way $p_{st}(x) = \lim_{s \rightarrow 0} s \hat{p}(x,s)$.
Taking the time derivative in Eq.\ (\ref{sup_main}) to zero and substituting Eq.\ (\ref{i_st}) we obtain the 
stationary advection-diffusion equation  
\begin{equation}
-a^{2}\nu \frac{d}{dx}\left[ F(x) p_{st}(x) \right] + \frac{d^2}{dx^2}\left[ D_{F}(x)p_{st}(x)\right] =0.
\label{stat_eq_2}
\end{equation}%
Integrating Eq.\ (\ref{stat_eq_2}) and taking into account that the flux of the particles is zero we obtain Eq. (\ref{stat_eq}).




\end{document}